\documentclass{article}
\usepackage{spconf,amsmath,graphicx}
\usepackage{amssymb}
\usepackage{mwe} 
\usepackage{booktabs}
\usepackage{amsfonts}

\title{Unsupervised Adaptation from FDG to PSMA PET/CT for 3D Lesion Detection under Label Shift}
 
\name{\parbox{\linewidth}{\centering Xiaofeng Liu, Menghua Xia, Yanis Chemli, Georges El~Fakhri, Chi Liu, Jinsong Ouyang}}
\address{Yale Biomedical Imaging Institute and Department of Radiology \& Biomedical Imaging, \\ Yale University, New Haven, CT 06520, USA}

\begin{document}
%
\maketitle

\begin{abstract}
In this work, we propose an unsupervised domain adaptation (UDA) framework for 3D volumetric lesion detection that adapts a detector trained on labeled FDG PET/CT to unlabeled PSMA PET/CT. Beyond covariate shift, cross tracer adaptation also exhibits label shift in both lesion size composition and the number of lesions per subject. We introduce self-training with two mechanisms that explicitly model and compensate for this label shift. First, we adaptively adjust the detection anchor shapes by re-estimating target domain box scales from selected pseudo labels and updating anchors with an exponential moving average. This increases positive anchor coverage for small PSMA lesions and stabilizes box regression. Second, instead of a fixed confidence threshold for pseudo-label selection, we allocate size bin-wise quotas according to the estimated target domain histogram over lesion volumes. The self-training alternates between supervised learning with prior-guided pseudo labeling on PSMA and supervised learning on labeled FDG. On AutoPET 2024, adapting from 501 labeled FDG studies to 369 $^{18}$F-PSMA studies, the proposed method improves both AP and FROC over the source-only baseline and conventional self-training without label-shift mitigation, indicating that modeling target lesion prevalence and size composition is an effective path to robust cross-tracer detection.
\end{abstract}


\vspace{-2pt}
\section{Introduction}\vspace{-2pt}
PET/CT imaging is important for cancer diagnosis and treatment planning, offering combined metabolic and anatomic information for detecting malignant lesions. Deep learning models have shown promise for automated 2D~\cite{weikert2023automated}, 2.5D~\cite{xu2023mask}, or 3D~\cite{chen2021multimodality,zhao2025coarse,liu2026aidriven} lesion detection in PET/CT. However, precisely annotating every lesion for a newly developed tracer can be costly, and a model trained on one tracer often fails to generalize to another due to domain shift~\cite{liu2022deep}. 

\begin{figure}[t]
\centering
\includegraphics[width=\columnwidth]{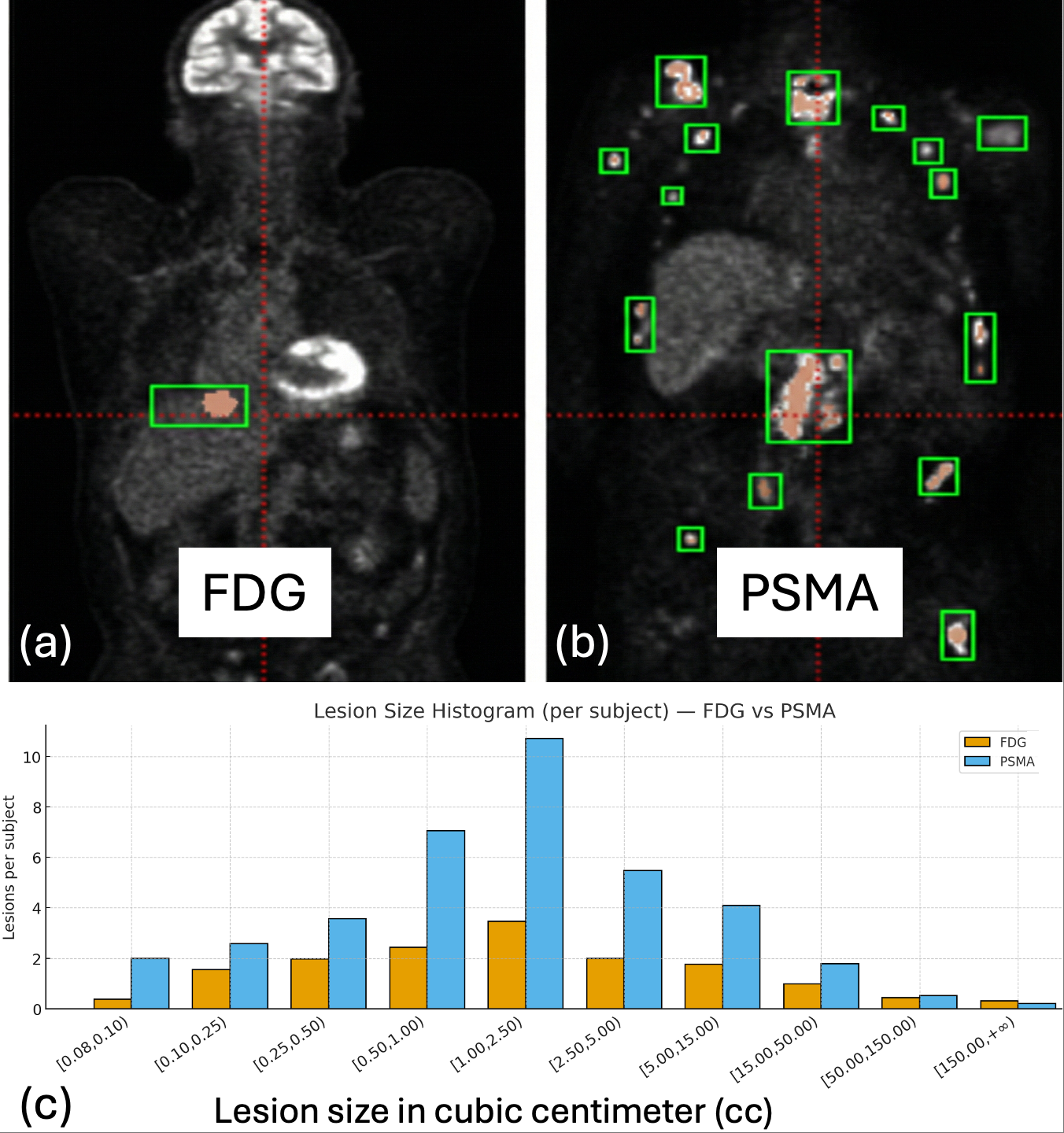}\vspace{-10pt}
\caption{(a-b) Examples of FDG and PSMA scans. (c) Lesion size distribution per subject for FDG vs. PSMA. On average, each PSMA subject contains more small lesions, whereas FDG has slightly more very large lesions ($\geq$150~cc). All volumes are resampled to $4\times4\times5$mm spacing.}\vspace{-5pt}
\label{fig:lesion-dist}
\end{figure}

The domain shift is particularly challenging for cross PET tracers applications. For example, a detection model trained on fluorodeoxyglucose (FDG) PET scans may underperform on prostate-specific membrane antigen (PSMA) PET scans because tracer uptake patterns and typical lesion characteristics differ significantly. Specifically, FDG highlights glucose-avid tumors, but also produces high physiological uptake in the brain and myocardium, while PSMA targets prostate cancer metastases with intense uptake in salivary glands and kidneys~\cite{gatidis2022whole}. Although co-scanned CT can provide the corresponding location information, the prevalence and size distribution of the lesions can vary between cohorts of patients imaged with these different tracers, leading to a possible \textit{label shift} in the frequency and characteristics of the lesions (Fig.~\ref{fig:lesion-dist}). Developing techniques to adapt models across such domains is critical to leverage knowledge from large scale labeled FDG datasets to new tracers like PSMA, where annotated examples can be scarce.

Unsupervised domain adaptation (UDA) methods can be a promising solution aimed at transfer of knowledge from a labeled source domain to a different unlabeled target domain~\cite{liu2022deep}. Recently, in the field of general computer vision, UDA has been applied to 2D image detection and 3D LiDAR point clouds detection~\cite{zhang2024cts}, but these have not been translated into 3D volumetric lesion detection.

More importantly, conventional UDA usually assumes that only the input appearance changes and that the label distribution remains consistent across domains. This assumption has led to a series of covariate shift UDA methods~\cite{liu2022deep}. However, the distribution of lesion detection bounding boxes differs markedly between FDG and PSMA PET/CT, which highlights the need to mitigate label shift together with appearance shift. Recently, the label shift has been explored in classification tasks~\cite{liu2021adversarial,liu2021domain}. However, to our knowledge, no prior work has addressed the label shift in UDA detection.

In this work, we propose a 3D UDA framework for lesion detection in whole-body PET/CT, adapting a RetinaNet detector from labeled FDG to unlabeled PSMA tracer. We use the multimodal 3D RetinaNet (M3DRetinaNet)~\cite{liu2026aidriven} as our detection backbone, which is a one-stage object detector with a focal loss to address sparse objects as well as  multi-scale pyramid features for diverse object sizes~\cite{lin2017focal}. Since each pyramid feature level has $\geq$1,000 anchor box proposals to be processed, it is not efficient to configure a discriminator at each level as adversarial UDA~\cite{liu2022deep}. Instead, we adapt the self-training UDA to generate the pseudo-label in target domain and alternating the supervised training in labeled source domain and pseudo-labeld target domains. We explicitly mitigate the label shift by incorporating prior knowledge on lesion size and prevalence, utilizing the target domain predictions to estimate the target label distribution. Then we are able to adjust the pseudo-parameter of anchor size used in M3DRetinaNet as well as the size bin-wise threshold for pseudo-label selection accordingly.

Our contributions can be summarized as follows:

\noindent$\bullet$ To our knowledge, this is the first attempt of unsupervised domain adaptation for 3D volumetric lesion detection.

\noindent$\bullet$ In addition to the conventional covariate shift, we design a novel self-training scheme with prior knowledge of label distribution shift, which has not been explored in UDA detection.

\noindent$\bullet$ The pseudo-parameter of size distributions and lesion prevalence in target domains are adaptively estimated to refine the pseudo-label selection.

We demonstrate its effectiveness on cross-tracer PET/CT (FDG to PSMA) lesion detection on AutoPET2024 dataset.
 
\vspace{-2pt}
\section{Methodology}\vspace{-2pt}
\label{sec:method}

Let the labeled source set be $\mathcal{D}^s=\{(x_i^s,y_i^s)\}_{i=1}^{N^s}$ from FDG PET/CT and the unlabeled target set be $\mathcal{D}^t=\{x_i^t\}_{i=1}^{N^t}$ from PSMA PET/CT. Each input is a dual-modality 3D volume $x\in\mathbb{R}^{2\times H\times W\times Z}$ formed by stacking PET and CT after a common resampling. The source label $y_i^s=\{(b_{ij},c_{ij})\}_{j=1}^{M_i}$ lists axis-aligned boxes $b_{ij}\in\mathbb{R}^6$ with one lesion class $c_{ij}=1$. The goal is to learn a detector $f_\theta$ that performs well on $\mathcal{D}^t$ without target training labels. Since the unknown target label distribution differs in lesion sizes and the number of lesions, we explicitly estimate the target domain priors of anchor-shape set $\mathcal{S}$ as well as size-density prior composed of a global mean $\mu$ (lesions per subject) and normalized histogram $h$ over $B$ volume bins.

\vspace{-5pt}
\subsection{3D detector and source pretraining}\vspace{-2pt}
We adopt the multimodal 3D RetinaNet (M3DRetinaNet)~\cite{liu2026aidriven} with a 3D ResNet backbone and a 3D feature pyramid network~\cite{baumgartner2021nndetection} operating on concatenated PET and CT channels. The detector relies on defining 3D anchor shapes $\{\mathbf{s}_k\}_{k=1}^K$. We apply the anchor boxes at each point in multiscale features, and the head predicts classification logits and box regressions to match ground truth box. With a mini-batch $\mathcal{B}$ and anchor index set $\mathcal{A}(\mathcal{B})$, the detection loss is
\begin{equation}
\label{eq:retinanet-loss}
\mathcal{L}_\mathrm{det}(\theta;\mathcal{B})=
\sum_{a\in\mathcal{A}(\mathcal{B})}\mathrm{FL}(p_a,c_a)
+ \!\!\sum_{a\in\mathcal{A}_+(\mathcal{B})}\ell_\mathrm{reg}(t_a,\hat{t}_a),
\end{equation}
where $\mathrm{FL}$ is focal loss~\cite{lin2017focal} for classification, $\ell_\mathrm{reg}$ is a smooth-L1-type regression loss, $p_a$ is the predicted lesion probability for anchor $a$, $c_a\in\{0,1\}$ is its assigned class, $t_a$ are predicted box parameters, and $\hat{t}_a$ are regression targets for positive anchors $\mathcal{A}_+$. Initial anchor shapes $\mathcal{S}^{(0)}=\{\mathbf{s}_k\}_{k=1}^K, K=3$ are obtained by $k$-means on source label boxes~\cite{liu2026aidriven,lin2017focal} to fit the source domain label distribution. We pretrain $f_\theta$ on $\mathcal{D}^s$ with 100 epochs to obtain a strong initialization.

\vspace{-2pt}
\subsection{Epoch-wise anchor size adaptation}\vspace{-2pt}
If anchor shapes are too different (e.g., large) relative to target domain lesions, the regression residual or error would be large. We therefore adapt the anchor shapes toward the target size statistics estimated from the selected pseudo labels. At epoch round $r$, let $\mathcal{B}^t_\mathrm{sel}=\bigcup_i \hat{y}_i^t$ be the union of selected pseudo boxes. We run $k$-means (in voxel space) on $\mathcal{B}^t_\mathrm{sel}$ to obtain centroids $\{\mathbf{s}_k^{(r)}\}_{k=1}^K$ and update the anchor set by an exponential moving average (EMA):
\begin{equation}
\label{eq:anchor-ema}
\mathbf{s}_k^{(r)} \leftarrow (1-\beta)\,\mathbf{s}_k^{(r-1)} + \beta\,\tilde{\mathbf{s}}_k^{(r)}, \quad k=1,\dots,K,
\end{equation}
where $\{\tilde{\mathbf{s}}_k^{(r)}\}_{k=1}^K$ are the $k$-means results on the selected target boxes, and $\beta\in(0,1)$. $\mathbf{s}_k^{(r)}$ is used for $r+1$ round predictions. 

\vspace{-2pt}
\subsection{Prior-guided pseudo-label selection}\vspace{-2pt}
Given the current detector, a forward pass on a target volume $x_i^t$ produces raw detections $\{(\hat{b}_{ij}, s_{ij})\}_j$, where $\hat{b}_{ij}\in\mathbb{R}^6$ is the box and $s_{ij}\in[0,1]$ its confidence. We first discard detections with $s_{ij}<\tau$ (we use $\tau=0.5$) and then apply 3D non-maximum suppression (NMS) with an IoU threshold of $0.25$ to remove largely overlapping boxes, resulting in a cleaned candidate set $\mathcal{C}_i$. The conventional fixed-threshold or top-$p\%$ selection~\cite{zou2019confidence,liu2022deep} would favor large, high-contrast lesions and bias the target toward source-like statistics. 

Instead, we select pseudo labels with more fine-grained size bin-wise threshold. Specifically, we use two online updated priors: a per subject budget controlling how many boxes a case should be selected, and a dataset-wise size-density histogram controlling which lesion sizes are retained. 

Let $\hat{y}_i^{t,(r)}$ denote the pseudo labels selected for subject $i$ in round $r$, and let $\widehat{M}_i^{(r)}=\lvert \hat{y}_i^{t,(r)}\rvert$. We maintain the global mean lesions-per-subject by an EMA:\vspace{-2pt}
\begin{equation}
\label{eq:mu-ema-sel}
\mu^{(r)} = (1-\alpha_\mu)\,\mu^{(r-1)} + \alpha_\mu\,\frac{1}{N^t}\sum_{i=1}^{N^t}\widehat{M}_i^{(r)},\vspace{-2pt}
\end{equation}
with momentum $\alpha_\mu\in(0,1)$. For the subsequent selection in round $r$, each subject has a per subject budget:\vspace{-2pt}
\begin{equation}
\label{eq:budget-sel}
N_{\mathrm{allow}}^{(r)}= \lambda\,\mu^{(r)},\vspace{-2pt}
\end{equation}
where $\lambda\in(0,1]$ controls how aggressively target detections should be admitted. This cap prevents a few scans with many false positives from dominating later rounds.

To encode the target size profile, we discretize lesion volume into $B$ bins and define a binning function $b:\mathbb{R}^6\to\{1,\dots,B\}$ that maps a box to its volume bin (in cc) under the working voxel spacing. From all pseudo labels selected at round $r$ we compute the normalized histogram $\widehat{h}^{(r)}$ and update the target size prior by EMA:
\begin{equation}
\label{eq:h-ema-sel}
h^{(r)}=\frac{(1-\alpha_h)\,h^{(r-1)}+\alpha_h\,\widehat{h}^{(r)}}
{\bigl\|(1-\alpha_h)\,h^{(r-1)}+\alpha_h\,\widehat{h}^{(r)}\bigr\|_1},\vspace{-2pt}
\end{equation}
with $\alpha_h\in(0,1)$ and $h^{(0)}$ initialized from the source histogram using identical bins.

During pseudo-label selection for a specific subject in round $r$, we allocate the budget $N_{\mathrm{allow}}^{(r)}$ to size bins according to $h^{(r)}$ by computing fractional quotas $\tilde{n}_b=h^{(r)}_b\,N_{\mathrm{allow}}^{(r)}$, taking integer parts $n_b=\lfloor \tilde{n}_b\rfloor$, and assigning any remaining slots to the bins with the largest fractional parts $\tilde{n}_b-n_b$ until $\sum_b n_b=N_{\mathrm{allow}}^{(r)}$. Each candidate $(\hat{b}_{ij}, s_{ij})\in\mathcal{C}_i$ is mapped to its bin $b_{ij}=b(\hat{b}_{ij})$. 

Within every bin we sort by $s_{ij}$ and retain the top-$n_b$ candidates. If a bin has fewer than $n_b$ candidates, we keep all of them. The final pseudo-label set for subject $i$ is\vspace{-2pt}
\begin{equation}
\label{eq:pseudo-y-sel}
\hat{y}_i^t=\{\,(\hat{b}_{ij},1)\,\}_{j\in\mathcal{S}_i^{(r)}},\vspace{-2pt}
\end{equation}
which is then used identically to ground-truth boxes in the detection loss in Eq.(\ref{eq:retinanet-loss}). By coupling a per-subject budget with bin-wise quotas driven by $h^{(r)}$, the selected pseudo labels track the evolving target lesion count and size composition rather than inheriting the FDG source bias.

\vspace{-2pt}
\subsection{Alternating optimization}\vspace{-2pt}
The mechanism integrates seamlessly with standard self-training~\cite{zou2019confidence,liu2022deep}. It changes only the selection rule for pseudo labels and the anchor shapes. We alternate supervised updates on source and pseudo-labeled target data while updating the priors and anchors:
\[
\textbf{Step 1:}\quad \min_{\theta}\,\mathbb{E}_{\mathcal{B}_s\sim\mathcal{D}^s}\big[\mathcal{L}_\mathrm{det}(\theta;\mathcal{B}_s)\big].
\]
\[
\textbf{Step 2:}\quad \text{Infer on }\mathcal{D}^t\Rightarrow\hat{y}_i^t;\ \text{update }\mu^{(r)},h^{(r)}\text{ by \eqref{eq:mu-ema-sel} and \eqref{eq:h-ema-sel};}
\]
\[
\text{update }\mathcal{S}^{(r)}\text{ by \eqref{eq:anchor-ema};}\quad
\min_{\theta}\,\mathbb{E}_{\mathcal{B}_t\sim\widehat{\mathcal{D}}^t}\big[\mathcal{L}_\mathrm{det}(\theta;\mathcal{B}_t)\big].
\]
We repeat for $R$ rounds. Each round comprises a full inference pass on $\mathcal{D}^t$, prior and anchor updates, and an epoch of supervised training on $\mathcal{D}^s$ and $\widehat{\mathcal{D}}^t$ sequentially.

It is worth noting that no adversarial loss or explicit feature alignment was used. The entire adaptation process is relatively lightweight and more efficient than trying to adversarially train multiple discriminators for four feature pyramid level each with dense ROI proposals. It can be unstable and impractically slow for 3D data. Our self-training approach leverages the solid initialization from FDG and gently nudges it toward the PSMA domain.

\begin{figure}[t]
\centering
\includegraphics[width=\columnwidth]{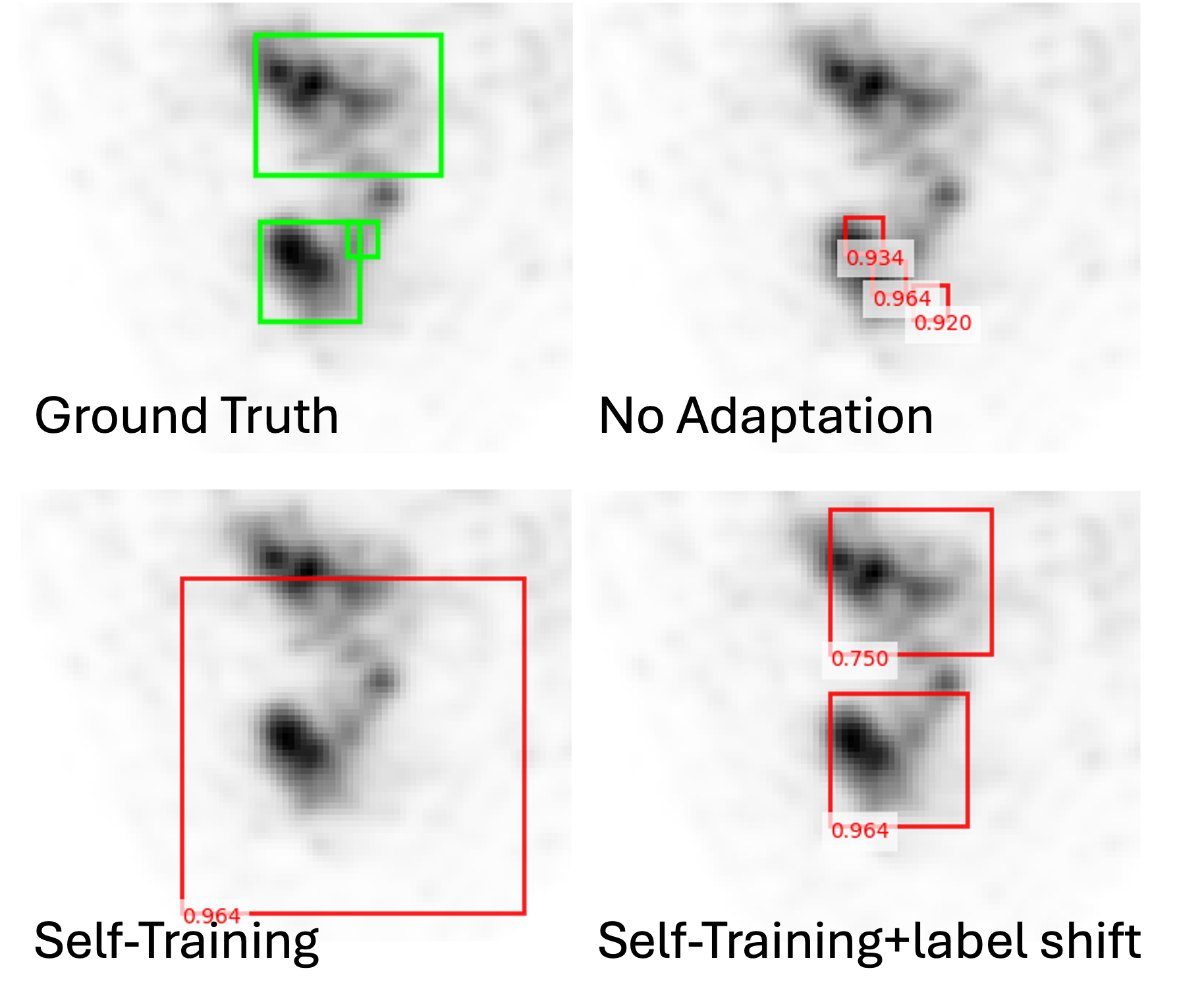}\vspace{-5pt}
\caption{Example lesion detection results on PSMA PET/CT.}
\label{fig:vis-examples}
\end{figure}

\vspace{-5pt}
\section{Experiments}\vspace{-2pt}

We evaluate on the public AutoPET2024 challenge dataset~\cite{gatidis2022whole}. We utilized 501 patient FDG PET/CT studies with lesion as the source domain and 369 $^{18}$F-PSMA PET/CT studies as the target domain. Specifically, 501 labeled FDG and 258 unlabeled PSMA studies (70\%) are used for UDA training, while 37 (10\%) and 74 (20\%) PSMA studies are used for target domain validation and testing, respectively. For the purpose of our detection task, we converted the segmentation masks into 3D bounding boxes that tightly encapsulate each contiguous lesion region. This provides ground-truth lesion boxes for training and evaluation. All scans are resampled to $4\times4\times5$mm spacing to match the resolution of some PSMA studies. PET SUV images and CT images were each z score normalized. Although the model can process an arbitrary size~\cite{baumgartner2021nndetection}, we randomly cropped a fixed 3D input size of $96\times96\times96$ voxels for efficient processing during training. The sliding window is applied for testing. For conventional self training with or without anchor adaptation, we select the top 50\% box predictions by confidence as pseudo labels. Notably, we do not need pseudo-label selection and its related hyperparameters in testing stage. Only the final $\mathcal{S}^{(r)}$ at the end of training is used as the anchor set for target domain testing. We empirically use $B{=}10$ volume bins (cc) covering the range in Fig.~\ref{fig:lesion-dist}. Hyperparameters $(\alpha_\mu,\alpha_h,\beta)$ are set to $0.9$ to stabilize early rounds. The selection factor $\lambda$ increases linearly from $0.1$ to $0.8$ over $R{=}200$ rounds. We discard lesions labels with volume $<0.08$ cc (less than one voxel at $4\times4\times5$ mm).

\begin{table}[t]
\centering
\caption{Lesion detection performance on PSMA PET/CT.}
\vspace{2pt}
\resizebox{\linewidth}{!}{%
\begin{tabular}{lccc}
\toprule
\textbf{Model} & AP@0.1 & AP@0.25 & AP@0.5  \\
\midrule
FDG source (no adapt)      & 0.225 & 0.184 & 0.128 \\
Ours (UDA adapted)         & 0.582 & 0.467 & 0.405 \\
Ours (UDA+anchor adapt)    &  {0.607} &  {0.472} &  {0.408} \\ 
Ours (UDA+label shift)     & \textbf{0.613} & \textbf{0.484} & \textbf{0.417} \\\hline
Fine-tuned on labeled PSMA         & 0.729 & 0.605 & 0.518 \\
\bottomrule
\end{tabular}%
}
\label{tab:results}
\end{table}


As shown in Fig.\ref{fig:vis-examples}, without self-training for covariate adaptation of the input appearance, the model tends to only predict on relatively high uptake regions and wrongly partitioned the lesion. Without modeling label shift, self-training tends to predict relatively larger and fewer lesions, following the label distribution in the FDG dataset. After we incorporate label shift, lesions can be detected more accurately.

Quantitatively, we evaluate lesion detection performance using both average precision (AP) and free-response receiver operating characteristic (FROC)~\cite{he2009roc} analysis, on a held-out set of target-domain (PSMA) images with ground truth labels. For AP, we treat lesion detection as a one-class object detection problem and compute the area under the precision-recall curve for detected lesions. We report AP at IoU thresholds of 0.1, 0.25, and 0.5 (IoU measured between predicted box and the nearest ground-truth lesion box). An IoU of 0.1 or 0.25 is fairly lenient, counting a detection as correct if it overlaps the lesion at least roughly, while 0.5 is more strict and evaluates localization accuracy. A detection is considered a true positive if its IoU $\ge$ threshold with any ground truth lesion and if each ground truth is matched to at most one detection (additional detections on the same lesion are false positives). For reference, we also evaluate the unadapted FDG model on the PSMA test set to quantify the domain shift impact, and an “upper-bound” model trained on FDG and fine-tuned with labeled PSMA training set for 200 epochs.  Table~\ref{tab:results} summarizes the detection performance on PSMA PET/CT. After adaptation, our model’s performance improves substantially. A simple anchor size adaptation can improve the performance, especially for AP@0.1. The full label shift mitigation with both anchor adapttaion and prior-guided pseudo-label selection achieved the best performance.

\begin{figure}[t]
\centering
\includegraphics[width=1\columnwidth]{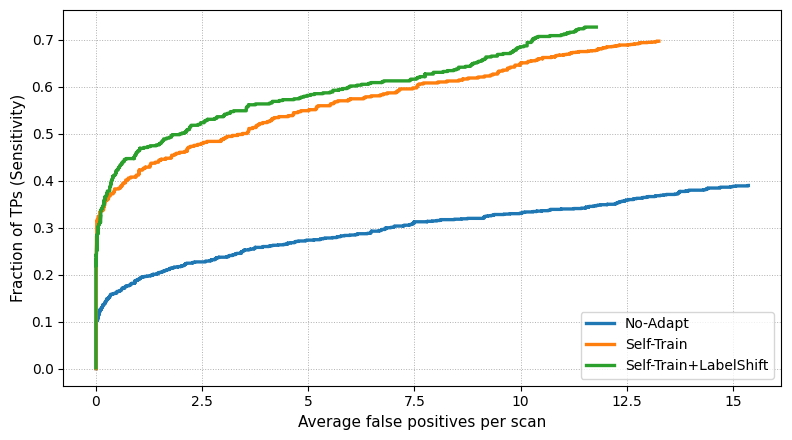}
\caption{Free-response ROC (FROC) curves on PSMA test set for the FDG-trained model (blue) and self-training UDA without (orange) or with (green) label shift correction.}
\label{fig:froc}
\end{figure}

FROC analysis provides a more comprehensive view of the sensitivity-versus-false-positive tradeoff. We plot the fraction of true positive lesions detected against the average number of false positives per scan, by sweeping the confidence cutoff from high to low. For our evaluation, a true positive is defined as a detection with IoU $\ge0.1$ with a ground-truth lesion, following prior works~\cite{liu2026aidriven}. The FROC curves in Fig.~\ref{fig:froc} illustrate that the self-training adapted detector achieves much higher sensitivity for any given false positive rate. Mitigating the label shift can further improve the performance.

\section{Discussion and Conclusion}
We presented a self-training framework for 3D lesion detection that explicitly addresses label shift when adapting from FDG to PSMA PET/CT. The method combines epoch-wise anchor shape adaptation with a target-side size-density prior that guides pseudo-label selection. It yields more accurate lesion localization and improves both AP and FROC on PSMA. The framework requires only model outputs on the target set, which is efficient for 3D data and can be added on to arbitrary detection base models. Limitations include reliance on accurate binning, which can under-represent highly clustered lesions, and the one-class setting, which simplifies label shift to size and count. Future work will consider uncertainty-weighted selection, adaptive binning driven by information criteria, and extension to multi-class settings.

\section{COMPLIANCE WITH ETHICAL STANDARDS} 
This research study was conducted retrospectively using human subject data made available in open access. Ethical approval was not required as confirmed by the license attached with the open-access data.

\section{ACKNOWLEDGMENTS} 
This work is supported in part by NIH grants R01CA275188, P41EB022544, R21EB034911, R01CA296305, R01CA290745, and NVIDIA Academic Grant Program.

 
\bibliographystyle{IEEEbib}
\bibliography{refs}

\end{document}